# Effect of UV light irradiation on charge neutralization in XPS measurements

Lei Zhu[a], Yunguo Yang[b], Jianhua Cai[a], Xuefeng Xu[a*], Liran Ma[c†] and Jianbin Luo[c]

[a]School of Technology, Beijing Forestry University, Beijing 100083, China.

[b]School of Mathematics and Physics, University of Science and Technology Beijing, Beijing 100083, China.

[c]State Key Laboratory of Tribology in Advanced Equipment, Tsinghua University, Beijing 100084, China.

**Abstract**

When XPS analyses are performed on insulator surfaces, shift and deformation of spectra peaks typically take place due to the surface charging. To achieve reliable XPS measurements, neutralization techniques have been widely adopted but their effectiveness are still limited, and thus, new neutralization technologies are urgently needed. Here, stable XPS spectra in which all the peaks undergo a reduced and nearly constant shift without significant deformation and broadening were obtained by introducing the UV light irradiation, implying that the introduction of the UV light can not only greatly attenuate the strength but also significantly improve both the temporal stability and the spatial uniformity of the surface charging during XPS measurements. This phenomenon, referred to as UV-assisted neutralization in this article, was found as effective as the most commonly used dual beam charge neutralization. Further observations show that the suppression of the charging issue comes from the adsorption of the UV-excited photoelectrons onto the X-ray irradiation region. This neutralization method, combined with the binding energy referencing, can be expected to become a promising alternative technique for solving the charging issues in XPS measurements.

---

[*] Corresponding author. E-mail: xuxuefeng@bjfu.edu.cn

[†] Corresponding author. E-mail: maliran@tsinghua.edu.cn

# 1. Introduction

X-ray photoelectron spectroscopy (XPS) can provide information about the chemical states, the bonding structure, and the elemental composition of the sample surface, and thus is currently one of the most widely used surface analytical techniques [1-6]. During XPS measurements, electrons (i.e., photoelectrons) are continuously excited from the surface region irradiated by the X-ray. For samples with poor conductivities, the lost electrons cannot be replenished timely and therefore accumulation of positive charges on the surface will occur. This phenomenon of losing charge neutrality on the surface is commonly called surface charging, which can shift the photoelectron peaks towards higher binding energy (BE) and even distort the shape of the peaks, making the information obtained from the spectra highly misleading [7-10]. In order to obtain accurate XPS spectra, sample surfaces should be adequately charge-neutralized during XPS measurements.

At Present, an electron flood gun or a combination of electron and ion beams is the most widely used method for charge neutralization [11-15]. Unfortunately, after neutralization, under- or over-compensation may often occur although a steady charging state may appear on the sample surface [16]. The imperfect neutralization leads to a nonzero electric potential on the sample surfaces which will shift the whole XPS spectrum along the binding energy axis [17]. To correct the spectrum, elements having characteristic peaks with known BEs, for example, the elements of the material itself [18], the widely existing adventitious carbon layer (AdC) [19-20], and the added known elements such as the implanted argon [21] or the sputter-deposited gold [22-24], have been used as the binding energy referencing. Among them, the BE scale referencing based on the C 1s peak of AdC has become part of both ASTM and ISO standards [25,26]. Besides, allowing the charges on the irradiated surface to find a pathway to the ground by heating [15,27-28] or by illumination [28,29-31] is also a feasible solution for charge minimization.

However, as the ISO standard indicates, there is no universally applicable method for charge neutralization in XPS measurements so far [26]. Current charge neutralization techniques cannot completely eliminate the charging issues, e.g., non-uniform charging is still insurmountable in some cases at present [7,24,32-34]. Heating or illumination is just helpful for a limited number of insulators who have appreciable pyroconductivity or photoconductivity. Therefore, to obtain an accurate spectrum, new neutralization technologies are urgently needed. In this paper, it is surprising to find that the addition of ultraviolet light can significantly mitigate the charging on bulk insulator surfaces irradiated by X-ray. The UV light irradiation can not only greatly attenuate the strength but also significantly improve both the temporal stability and the spatial uniformity of the surface charging during XPS measurements. Then, the effectiveness of such a charge neutralization has been studied in detail and been compared with the standard dual beam charge neutralization. Furthermore, the neutralization mechanisms of UV light irradiation has been explored and the results show that the neutralization effect can be attributed to the adsorption of UV-excited photoelectrons onto the X-ray irradiation region.

## 2. Methods

The XPS measurements discussed in this paper were performed by using photoelectron spectroscopy (PES, PHI-5000 VersaProbe III, JAPAN). The monochromatic Al Kα X-ray line (hν=1486.6 eV) with a line width of 0.85 eV was employed as excitation source and the He-I line (hν=21.2 eV) with the lamp power of 65 W and the inherent energy resolution of about 0.1 eV was employed as the UV irradiation source. The base pressure in the ultra-high vacuum (UHV) chamber during experiments was about $10^{-8}$ mbar. A Spherical Capacitor Analyzer was used to analyze the energy of the photoelectrons and all the binding energies of the obtained spectra were referenced to the Fermi level (FL) of the spectrometer. The UPS spectra obtained here were performed with a pass energy of 2.6 eV and a step size of 0.02 eV.

The samples used here are three kinds of bulk insulators, i.e., round disks of α-$Al_2O_3$ crystal (0001) and of $SiO_2$ glass with a diameter of 30 mm and a thickness of 2 mm, and square disk of polyethylene terephthalate (PET) with a side length of 25 mm and a thickness of 0.5 mm. Fresh PET slices were used in this study to avoid aging or contamination as much as possible. All the samples were adhered to the sample holder by double-sided carbon tape. The standard dual beam charge neutralization, which utilizes simultaneously both a cold cathode electron flood source with an emission current of 20 µA and a very low energy ion source, was used as a comparison to estimate the effectiveness of the neutralization performance of the UV irradiation. In XPS measurements, the angle between the sample surface normal and the entrance of the input lens of the analyzer, i.e., the take-off angle (TOA), was set at 90° for XPS measurements with UV irradiation and 45° for XPS measurements with the dual beam charge neutralization.

## 3. Results and Discussion

### 3.1 UV-assisted charge neutralization

To demonstrate the charging caused by the focused monochromatic X-ray, a 100 µm X-ray beam with a working voltage of 15 kV and a power of 50 W irradiates continuously the surface of α-$Al_2O_3$ crystal (0001) and two spectra within binding energy ranges of 128-148 eV and 583-603 eV, which contain Al 2p peak and O 1s peak respectively, were measured sequentially and repeatedly without any charge neutralization. As shown in Figures 1a and 1b, the accumulation of positive charges on the sample surface caused by the irradiation leads to shifts of the peaks of approximately 60 to 70 eV towards higher binding energy. After 40 rounds of measurements, not only the positions of but also the distance between the two peaks still exhibits obvious and irregular fluctuations, indicating that the surface charging has not yet reached stability. In addition, the significant differences in peak shape among different measurements as shown in Fig. 1c and 1d show remarkable non-uniform surface charging (i.e., the differential charging [7]) across the irradiated surface region.

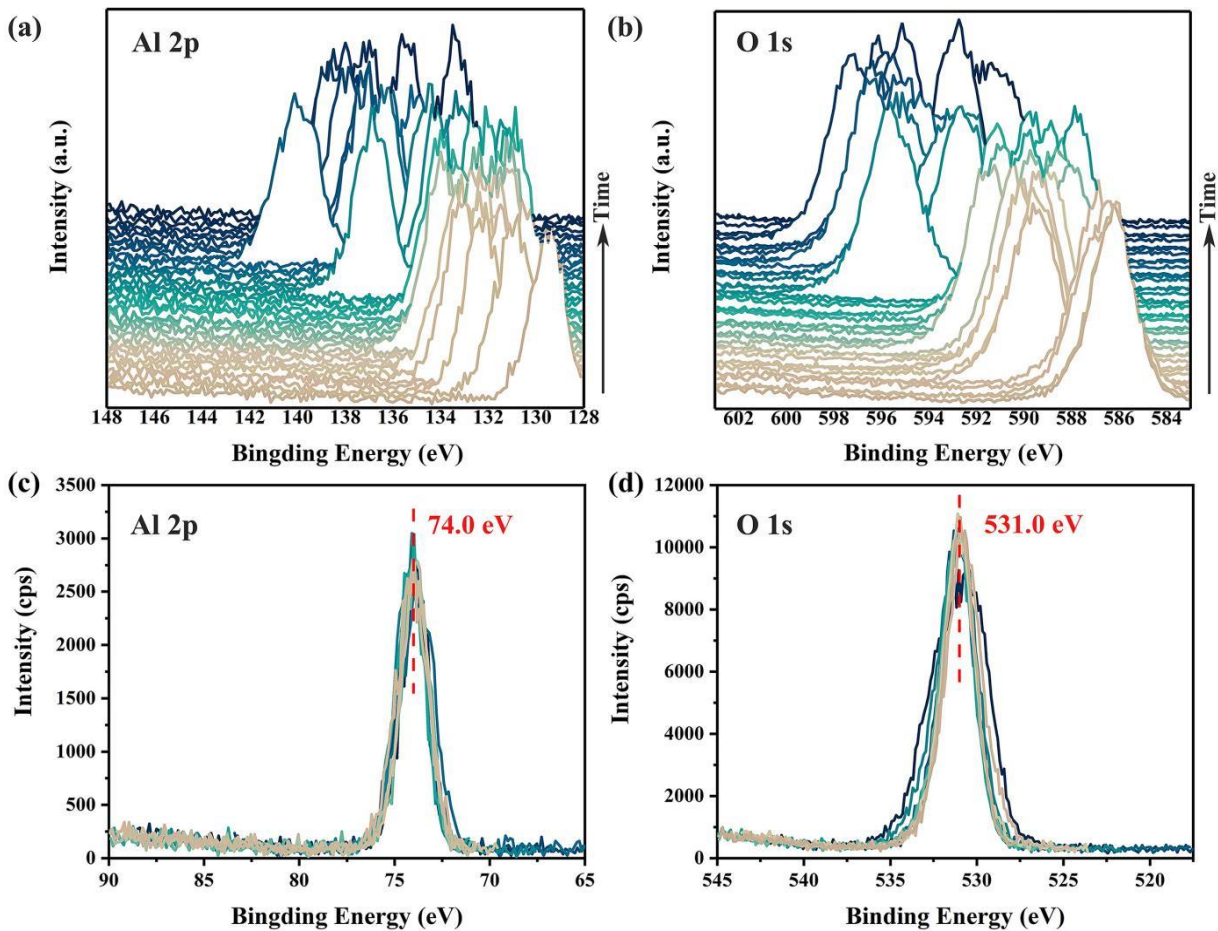

**Figure 1**. The Al 2p spectra (a) and the O 1s spectra (b) showing time-dependent charging on irradiated α-Al$_2$O$_3$ surface. The differences in peak shape of Al 2p (c) and O 1s (d) among different measurements showing the differential charging across the irradiated surface. The spectra were acquired using a pass energy of 69 eV and a scanning step of 0.125eV, with a total time spent of about 45 minutes.

Similar to X-ray irradiation, ultraviolet light irradiation can also cause the accumulation of positive charges on the irradiated surface and lead to a binding energy shift [35-37]. Thus, one can easily draw a conclusion based on intuition that the charging effect should be strengthened when both of the two lights irradiate on the same surface region. Surprisingly, it is found here that the addition of ultraviolet light irradiation can dramatically mitigate the charging on the X-ray irradiated surface (see Fig. 2). When neither the UV irradiation nor the dual beam charge neutralization is introduced, the charging-induced binding energy shift in XPS spectra of 10 consecutive measurements on α-Al$_2$O$_3$ crystal are time-dependent, varying between 70 to 100 eV (see Fig. 2a). The charging on SiO$_2$ glass surface or on polyethylene terephthalate (PET) surface is even more severe, making it impossible to obtain an effective

XPS spectra (see Fig. 2b and 2c). After the UV light irradiates on the sample surfaces, the peak shifts in the XPS spectra and their fluctuation throughout 10 consecutive measurements have been reduced to 22 eV and 0.11 eV, 18.5 eV and 0.06 eV and 17 eV and 0.12 eV for α-Al$_2$O$_3$, SiO$_2$ glass, and PET, respectively (see Fig. 2). This indicates that the irradiation of the UV light can not only greatly attenuate the strength but also significantly increase the stability of the charging on the sample surface caused by the X-ray irradiation. Such a charge neutralization method is called here the UV-assisted neutralization.

## 3.2 Effectiveness of UV-assisted neutralization

In order to evaluate the performance of the UV-assisted neutralization, narrow spectra of C 1s peaks of PET, an insulating polymer with well-established XPS spectra [15,17,38], were continuously measured 20 times with a pass energy of 26 eV and a step size of 0.05 eV in the presence of the UV light. The results show that the position, the shape, and the intensity of the C 1s peaks remain almost the same in all the measurements (see Fig. 3a). The standard deviation of the fluctuation of the charging-induced surface potential was calculated from the positions of the C-C peak and was found not more than 0.03 V, implying that the charging on the sample surface should be very stable. Further, the Full Widths at Half Maximum (FWHM) of the O-C=O peak, which is not easily affected by the structure variations of PET [38], was found to be about 0.85 eV (see Fig. 3b) which is equal to the line-width of X-ray light source, implying a nearly homogeneous charging across the measured region. These mean that UV light irradiation can improve significantly not only the temporal stability but also the spatial uniformity of the surface charging in XPS.

Although the present UV-assisted neutralization is under-neutralized and thus has a higher charging-induced surface potential compared with the dual beam charge neutralization (see Fig. 3c), the normalized spectra obtained by these two methods are almost identical (see Fig. 3d). Thus, it can be reasonably concluded that the UV-assisted neutralization is at least as good as, if not superior to, the dual beam charge neutralization, and thus may provide a new feasible way for solving charging issues in XPS.

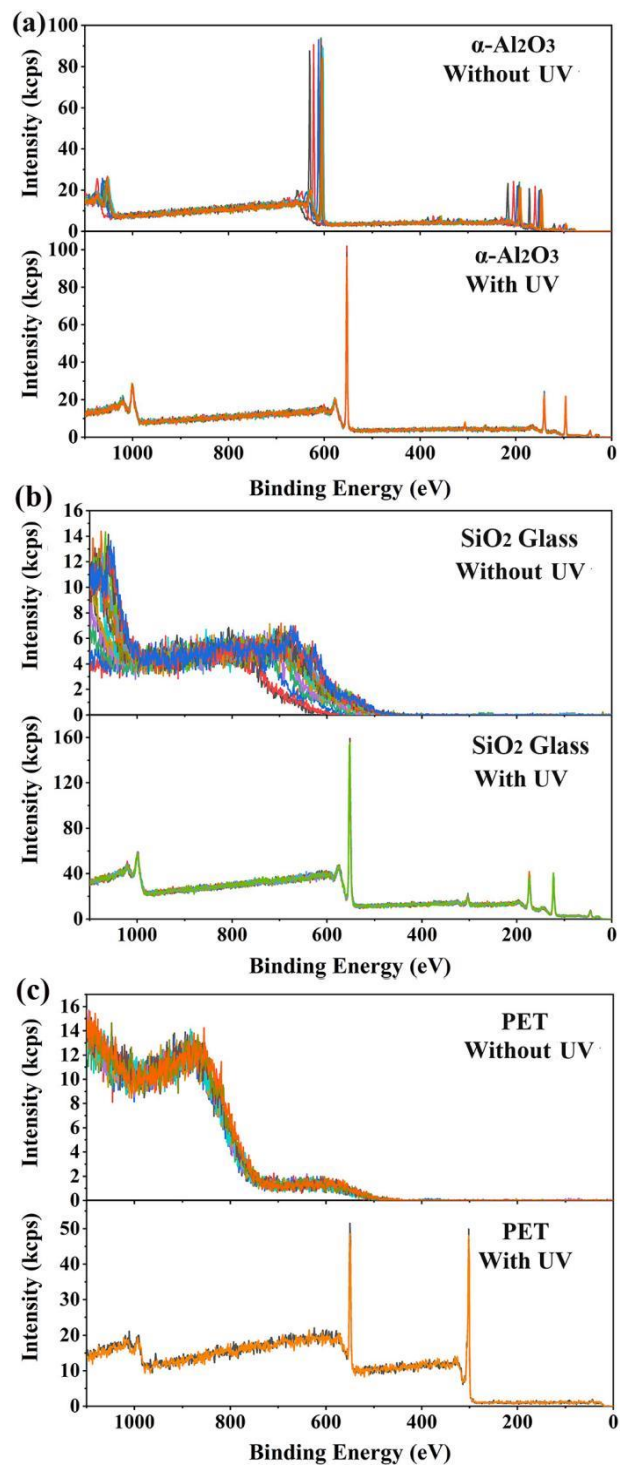

**Figure 2**. 10 consecutive measurements of full XPS spectra on (a) α-Al$_2$O$_3$ (0001) surface, (b) SiO$_2$ glass surface, and (c) PET surface without and with the UV light irradiation. In all the measurements, the dual beam charge neutralization was not used and the spectra were acquired with a pass energy of 280 eV and a scanning step of 1 eV.

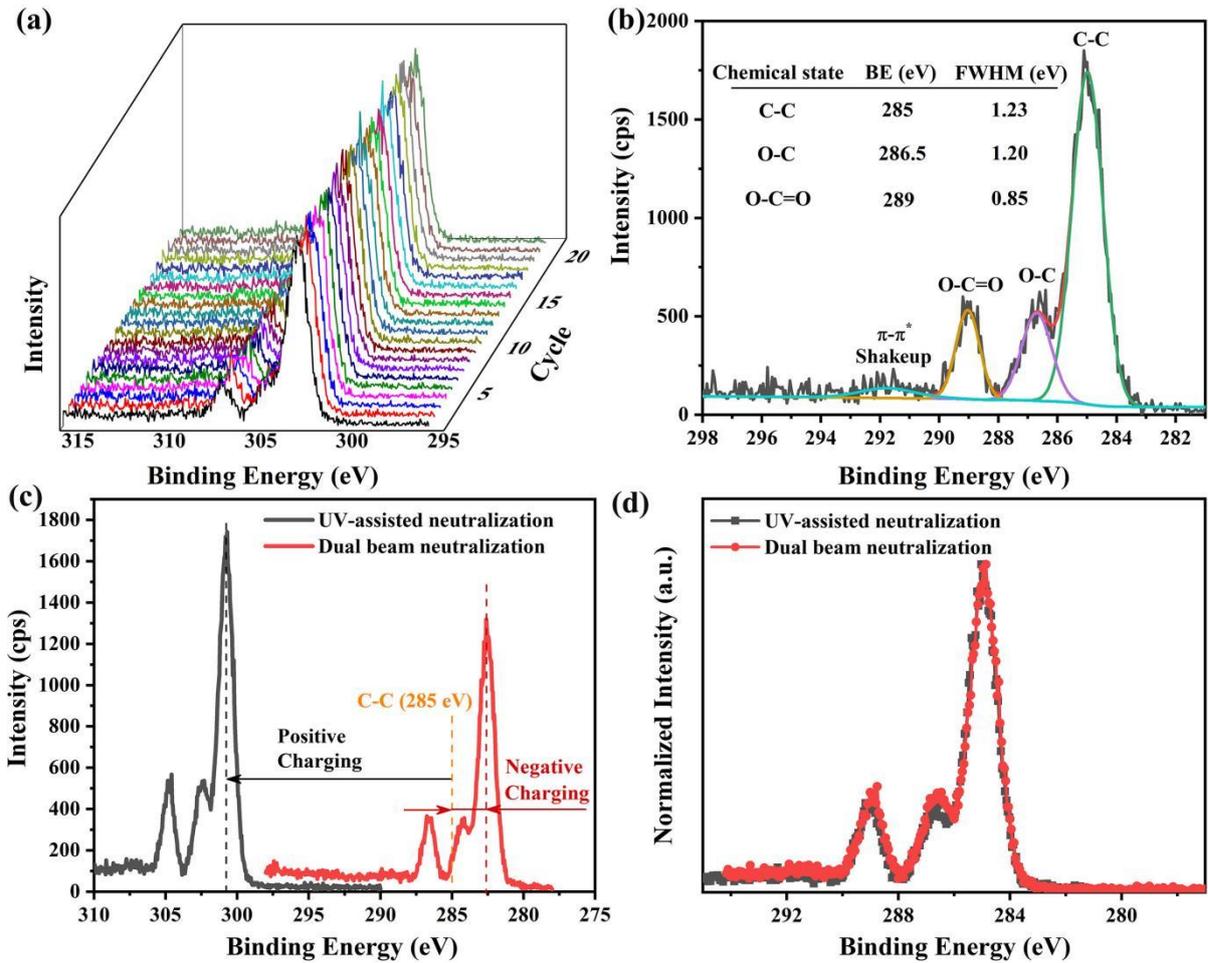

**Figure 3**. The neutralization performance of UV irradiation on PET surface. (a) 20 consecutive measurements on the narrow spectra of C 1s peaks on PET surface with UV-assisted neutralization. (b) The charge-corrected C 1s spectrum measured with UV-assisted neutralization and the peak-fitted results. (c) Spectra of C 1s peaks measured with the UV-assisted neutralization and that with the dual beam charge neutralization. (d) Comparison of the corrected and normalized C 1s peaks between the two neutralization methods. A 100 μm X-ray with a beam power of 50 W was used, and all spectra were collected using a pass energy of 26 eV and a step size of 0.05 eV.

### 3.3 Neutralization mechanisms of UV light irradiation

Our interesting findings imply that the electron loss on the sample surface caused by the X-ray irradiation can be compensated to some extent due to the introduction of the UV light irradiation. The possible increase in the surface conductivity induced by UV irradiation [29] is unlikely to be the cause of the charge neutralization because the size and thickness of the samples used here are much larger than the spot size and the penetration depth of the lights in PES so that the accumulated charges on the irradiation surface cannot find a pathway to the ground. To substantiate this point, the current from the sample holder to the ground during the XPS measurements with UV-assisted neutralization was measured by a Keithley 6517B electrometer and was found consistent with the background noise (see Fig. 4a).

To explore the neutralization mechanisms of UV light irradiation, XPS measurements in the presence of the UV light were performed on different points 1# - 7# arranged sequentially from the middle to the edge of the PET surface as shown in Fig. 4b. For the UV beam with a diameter of about 3mm, all the irradiation areas of the UV beam are located within the sample surface at points 1#-3#, but from points 4#-7#, as the measurement point is approaching the sample edge, more and more UV light will irradiate on the surface of the metal stage. The measurements show that the UPS spectra are almost the same at points 1# - 3#, but undergo significant changes in both the number and the energy distribution of the photoelectrons from point 4# to 7#, illustrating that a larger irradiation area on the metal stage will result in a higher yield of UV-excited photoelectrons (see Fig. 4c).

Due to the small diameter of the X-ray beam (100 μm), the irradiation area of the X-ray beam at all the measurement points will be located within the sample surface, and thus the XPS spectra at all the measurement points have almost the same shape (see Fig. 4d). However, the energy shifts along the binding energy axis, which represent the surface potential on the X-ray irradiation areas, are basically the same in the central region (i.e., points 1# - 3#), and gradually decrease as the point approaches the sample edge (from point 4# to point 7#). The decrease in the surface potential with the increase of the UV-excited photoelectrons means that the photoelectrons induced by UV irradiation are likely to be the cause of the charge neutralization.

The charging on the insulator surfaces irradiated by UV light can reach its steady state very quickly (within 1 s) and both the magnitude and the fluctuation of the charging-induced surface potential are very small (about 2 V and within 0.02 V respectively) [35]. When electrons are excited from the UV-irradiated surface region, the electrons (i.e., the compensation electrons) with kinetic energy lower than the electric potential energy of the X-ray irradiation surface can be attracted onto the surface by the electric force and neutralize the charges there. By integrating the compensation electrons in the UPS spectrum (see the shaded regions in Fig. 4c), the compensation current can be obtained. The strong dependence of the surface potential of the X-ray irradiated surface on the compensation current as illustrated in Fig. 4e clearly shows that the adsorption of the UV-excited photoelectrons onto the X-ray irradiation region should be the primary mechanism of the UV-assisted neutralization.

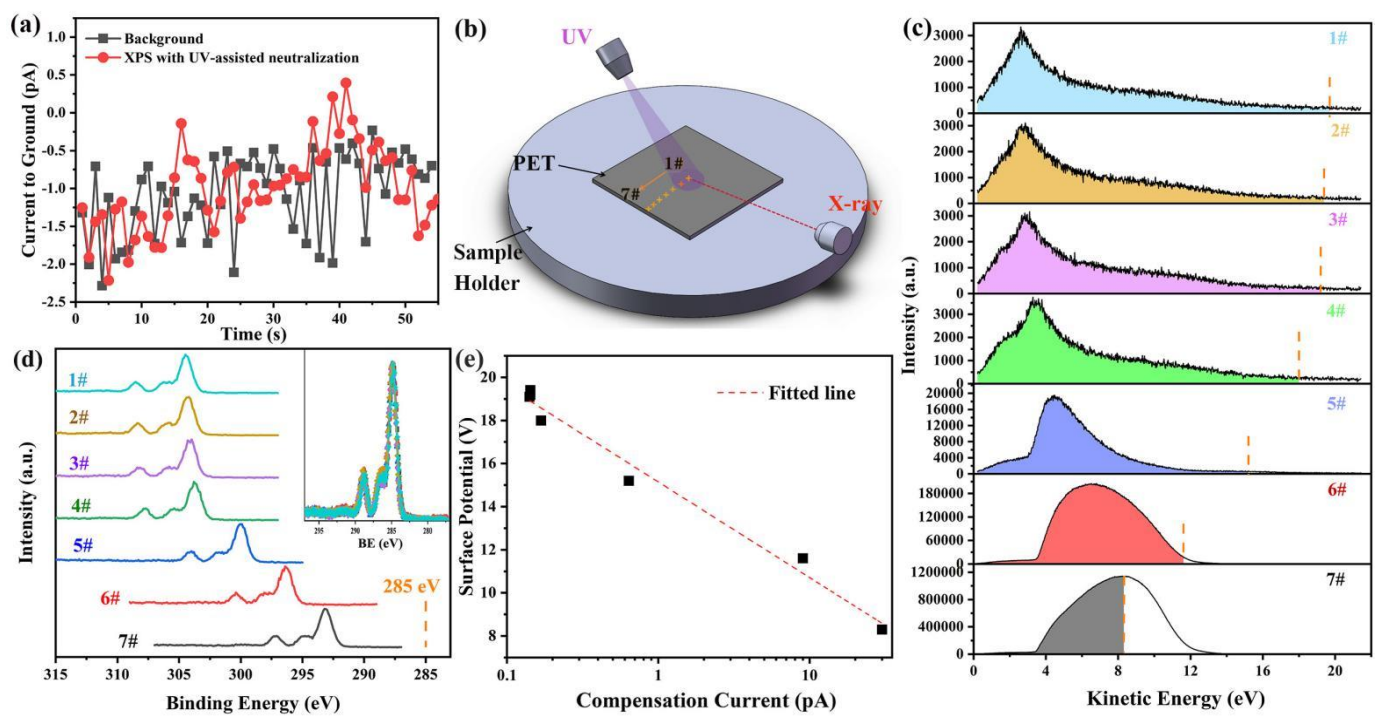

**Figure 4**. (a) The current to the ground of the sample holder during XPS measurements with UV-assisted neutralization and its comparison with the background noise. (b) Schematic diagram of XPS measurements with UV-assisted neutralization on different points 1# - 7# arranged sequentially from the middle to the edge of the PET surface. (c) The UPS spectra obtained at different measurement points in the absence of X-ray. The vertical dashed lines correspond to the surface potential on the X-ray irradiation region in the corresponding measurement and the shaded regions represent the compensation electrons. (d) The narrow spectra of C 1s peaks obtained at different points with UV irradiation, and the inset shows these spectra after charge correction and intensity normalization. (e) The relationship between the surface potentials on the X-ray irradiated surface region and the compensation currents.

## 4. Conclusion

This study introduces a novel charge neutralization technique called UV-assisted neutralization. For XPS measurements on the surface of bulk insulator samples, it is found that ultraviolet light irradiation can dramatically mitigate the charging issues in the X-ray irradiation region, greatly reducing the charging-induced surface potential and significantly improving both the temporal stability and the spatial uniformity of the surface charging. The performance of the UV-assisted neutralization in XPS measurements on PET surface is found to be at least as good as, if not superior to, that of the standard dual beam charge neutralization.

The strong dependence of the surface potential of the X-ray irradiated surface on the yield of UV-excited photoelectrons clearly shows that the absorption of UV-excited photoelectrons onto the X-ray irradiation region is likely the primary mechanism of the UV-assisted neutralization. It can be reasonably expected that the neutralization effect may be further improved if the yield of UV-excited photoelectrons is increased. Although it cannot completely eliminate the charging issues in XPS, e.g., under-compensation may occur, the UV-assisted neutralization can provide a very stable and uniform charging surface, and thus may become a feasible alternative solution to the charging problem in XPS.